\documentclass[conference]{IEEEtran}
\IEEEoverridecommandlockouts

\usepackage{cite}
\usepackage{amsmath,amssymb,amsfonts}
\usepackage{algorithmic}
\usepackage{graphicx}
\usepackage{textcomp}
\usepackage{xcolor}
\usepackage{kotex}
\usepackage{mathtools}
\def\BibTeX{{\rm B\kern-.05em{\sc i\kern-.025em b}\kern-.08em
    T\kern-.1667em\lower.7ex\hbox{E}\kern-.125emX}}
\begin{document}

\newcommand{\bumsu}[1]{\color{blue}{[BUMSU, #1]}\color{black}}
\newcommand{\heedong}[1]{\noindent{\textsf{[Heedong]: {\color{red}#1}}}}
\newcommand{\brm}[1]{\boldsymbol{\mathrm{#1}}}
\newcommand{\bs}[1]{\boldsymbol{#1}}

\title{Transformer-Based Nonlinear Transform Coding for Multi-Rate CSI Compression in MIMO-OFDM Systems}

\author{\IEEEauthorblockN{Bumsu Park}
\IEEEauthorblockA{\textit{School of Electrical Engineering} \\
\textit{Korea University}\\
Seoul, South Korea\\
bumsupark@korea.ac.kr}
\and
\IEEEauthorblockN{ Heedong Do}
\IEEEauthorblockA{\textit{School of Electrical Engineering} \\
\textit{Korea University}\\
Seoul, South Korea\\
doheedong@korea.ac.kr}
\and
\IEEEauthorblockN{Namyoon Lee}
\IEEEauthorblockA{\textit{Department of Electrical Engineering} \\
\textit{POSTECH}\\
Pohang, South Korea\\
nylee@postech.ac.kr}
}

\newtheorem{theorem}{Theorem}
\newtheorem{lemma}[theorem]{Lemma}
\maketitle

\begin{abstract}
We propose a novel approach for channel state information (CSI) compression in multiple-input multiple-output orthogonal frequency division multiplexing (MIMO-OFDM) systems, where the frequency-domain channel matrix is treated as a high-dimensional complex-valued image. Our method leverages transformer-based nonlinear transform coding (NTC), an advanced deep-learning-driven image compression technique that generates a highly compact binary representation of the CSI. Unlike conventional autoencoder-based CSI compression, NTC optimizes a nonlinear mapping to produce a latent vector while simultaneously estimating its probability distribution for efficient entropy coding. By exploiting the statistical independence of latent vector entries, we integrate a transformer-based deep neural network with a scalar nested-lattice uniform quantization scheme, enabling low-complexity, multi-rate CSI feedback that dynamically adapts to varying feedback channel conditions. The proposed multi-rate CSI compression scheme achieves state-of-the-art rate-distortion performance, outperforming existing techniques with the same number of neural network parameters. Simulation results further demonstrate that our approach provides a superior rate-distortion trade-off, requiring only 6\% of the neural network parameters compared to existing methods, making it highly efficient for practical deployment.

\end{abstract}


\section{Introduction}

In massive multiple-input multiple-output (MIMO) systems, having accurate channel state information at the transmitter (CSIT) is crucial for achieving effective beamforming and spatial multiplexing gains \cite{marzetta2010noncooperative}. However, realizing these benefits comes with significant challenges. At high signal-to-noise ratios, where multiplexing gains are most advantageous, the effectiveness of beamforming and multiplexing techniques becomes highly dependent on the accuracy of CSIT \cite{GPIP,GPIP2,GPIP3}. Obtaining precise CSIT is particularly difficult in frequency-division duplex (FDD) systems, where users must measure the channel state and relay this information back to the transmitter. This challenge grows as the number of antennas at the base station (BS) increases. For example, in 6G networks, the BS is expected to utilize several hundred antennas, which further complicates the situation \cite{Lin_Lee_book,Tat20216G}.

CSI compression in massive MIMO systems is crucial for reducing the CSI feedback overhead while preserving high CSIT accuracy. In these systems, CSI is typically represented as a complex-valued vector, where each element corresponds to the channel gain and phase shift between a pair of transmit and receive antennas. Traditional CSI compression methods rely on codebook-based vector quantization techniques \cite{love2008overview}. In these approaches, the BS and the user share a predefined set of quantized channel representations, known as a codebook, before communication. Each codebook entry corresponds to a quantized approximation of the channel vector. To compress the high-dimensional channel vector, the user selects the codebook entry that best matches the channel and encodes it as an index pointing to that entry. Codebooks are generally designed to capture common variations in channel conditions and can be optimized based on criteria such as minimizing quantization error or maximizing system throughput. Grassmannian codebooks \cite{Love2003Grass}, which optimize vector alignment in multiuser scenarios, are commonly used in MIMO systems. Oversampled discrete Fourier transform matrices are also widely adopted in commercial wireless standards, including 3GPP \cite{3gpp_ts_38_211}. However, there is a major limitation of these approaches that the codebook size must scale exponentially with the number of antennas to keep the quantization error within acceptable bounds \cite{jindal2004limited,Lee2011}. It is worth noting that there has been some research that utilized partial reciprocity of channel parameters such as angle and delay \cite{Jungyeon2025,GPIP4}. These branches of study performed beamforming with minimal overhead \cite{GPIP4}, even without CSI feedback \cite{Jungyeon2025}.

\begin{figure*}[tbp]
\centerline{\includegraphics[width=0.85\textwidth]{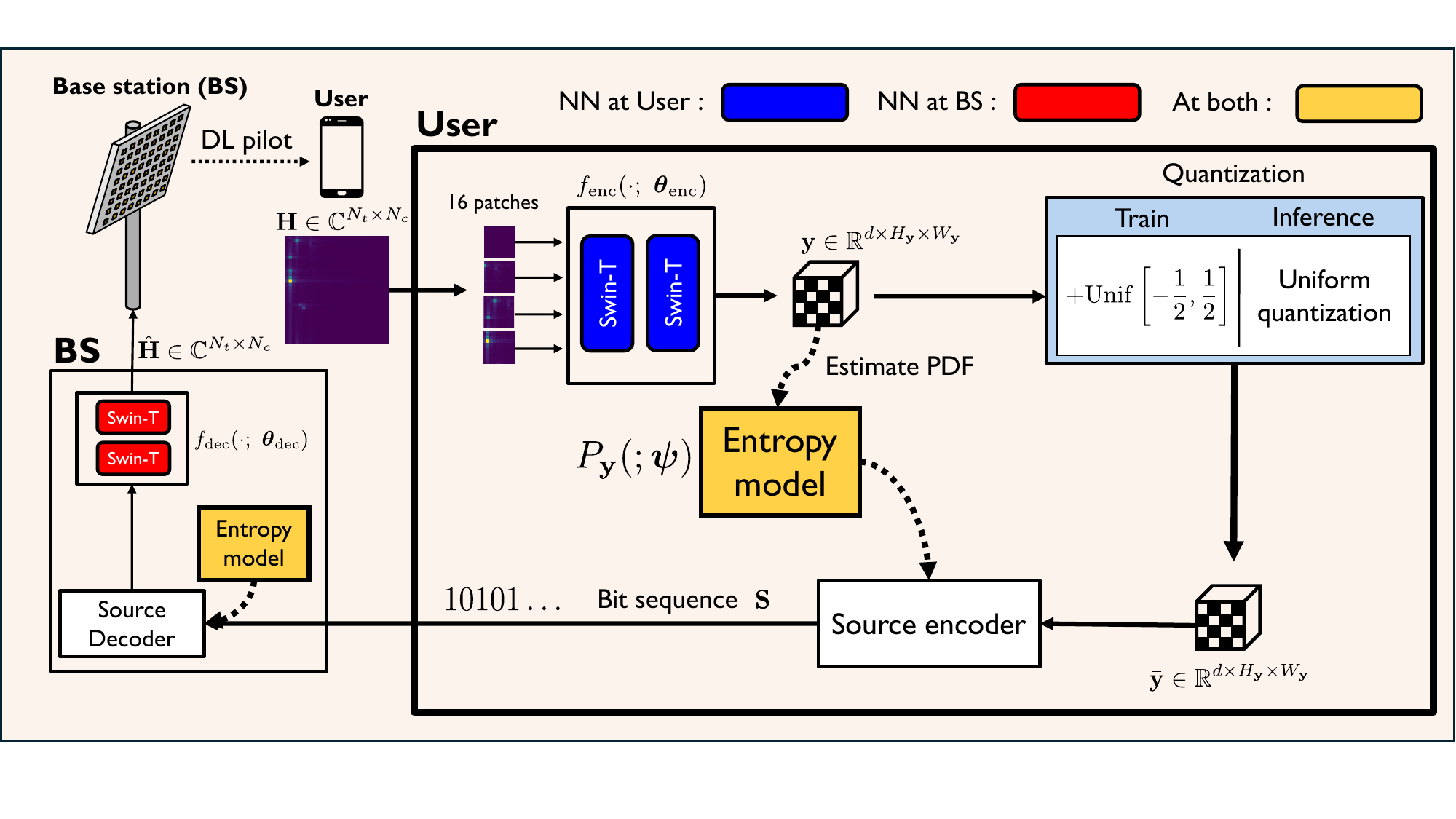}}
\caption{Block diagram of the bit-wise CSI feedback framework using NTC.}
\label{fig1:ntc}
\end{figure*}

Recently, deep learning (DL)-based CSI compression methods have gained significant attention due to the DL's advantages in efficiently compressing complex data. These techniques typically utilize deep autoencoder models, which consist of jointly optimized encoder and decoder networks. At the user side, the encoder compresses the estimated channel into a compact latent vector, which is then transmitted to the BS. The BS uses the decoder to reconstruct the full channel vector. For example, CsiNet employs a convolutional neural network (CNN)-based autoencoder to reduce the dimensionality of the estimated channel vector. To adaptively control CSI feedback rates based on feedback channel conditions, recent DL-based CSI compression methods have incorporated multi-rate quantization techniques alongside CNN-based autoencoders \cite{bi2022novel, mashhadi2020distributed, bumsu2024multirate}.

In this paper, we introduce an innovative method for compressing CSI in massive MIMO systems that use orthogonal frequency division multiplexing (OFDM). In these systems, the frequency-space channel matrix can be viewed as a high-dimensional, two-channel image, where each channel represents the real and imaginary parts of the complex-valued channel matrix. 

The proposed approach utilizes a cutting-edge image compression technique called transformer-based nonlinear transform coding (NTC) \cite{balle2017end}. NTC is an effective image compression approach that employs deep neural networks to convert an image into a compact binary format. Unlike traditional deep learning-based CSI compression methods, NTC simultaneously learns a nonlinear mapping to create the latent vector and the latent distribution for entropy coding. By taking advantage of the statistical quasi-independence among the entries in the latent vector, we integrate a transformer-based deep neural network with a scalar quantization scheme that features a nested-lattice structure along a real line. This combination allows for adaptive multi-rate CSI feedback based on the conditions of the feedback channel.

Our main finding is that this multi-rate CSI compression method delivers state-of-the-art rate-distortion performance when compared to existing techniques, all while using the same number of neural network parameters. Additionally, simulations using the COST2100 indoor dataset show that our approach achieves a significantly better rate-distortion trade-off with only 6\% of the neural network parameters used in current methods.




\section{System Model}
This paper considers a FDD massive MIMO system where BS uses $N_t$ number of antennas and the user has a single antenna. In addition, we also consider OFDM waveform using $\tilde{N}_c$ sub-carriers. Altogether, the downlink channel is defined as ${\bf H}\in \mathbb{C}^{\tilde{N}_c\times N_t}$. In this paper, we assume that the user perfectly estimates the downlink channel via downlink channel estimation. The user compresses ${\bf H}$ into a bit sequence $\brm{s}$ through preprocessing, DL-based encoding, and entropy encoding. We assume that the bit sequence is transmitted to the BS using channel coding and the BS recovers the CSI using an entropy decoder and a DL-based decoder.

\subsection{Preprocessing}
A channel that experiences long delays loses much of its power. As a result, the parts of the channel's time-domain signal that are delayed beyond a certain threshold become nearly zero and carry no useful information. If we define this threshold for near-zero delays as $N_c$, and both the BS and the user agree on it, the user can minimize feedback by only sending information about the channel dimensions with delays less than or equal to $N_c$, resulting in minimal information loss. Consequently, similar to previous studies \cite{wen2018deep, liang2022changeable}, all channels of ${\bf H}$ that the user reports are treated as dimensionally compressed channels after this preprocessing step, and ${\bf H}$ is redefined as the preprocessed channel.

\begin{figure*}[tbp]
\centerline{\includegraphics[width=0.8\textwidth]{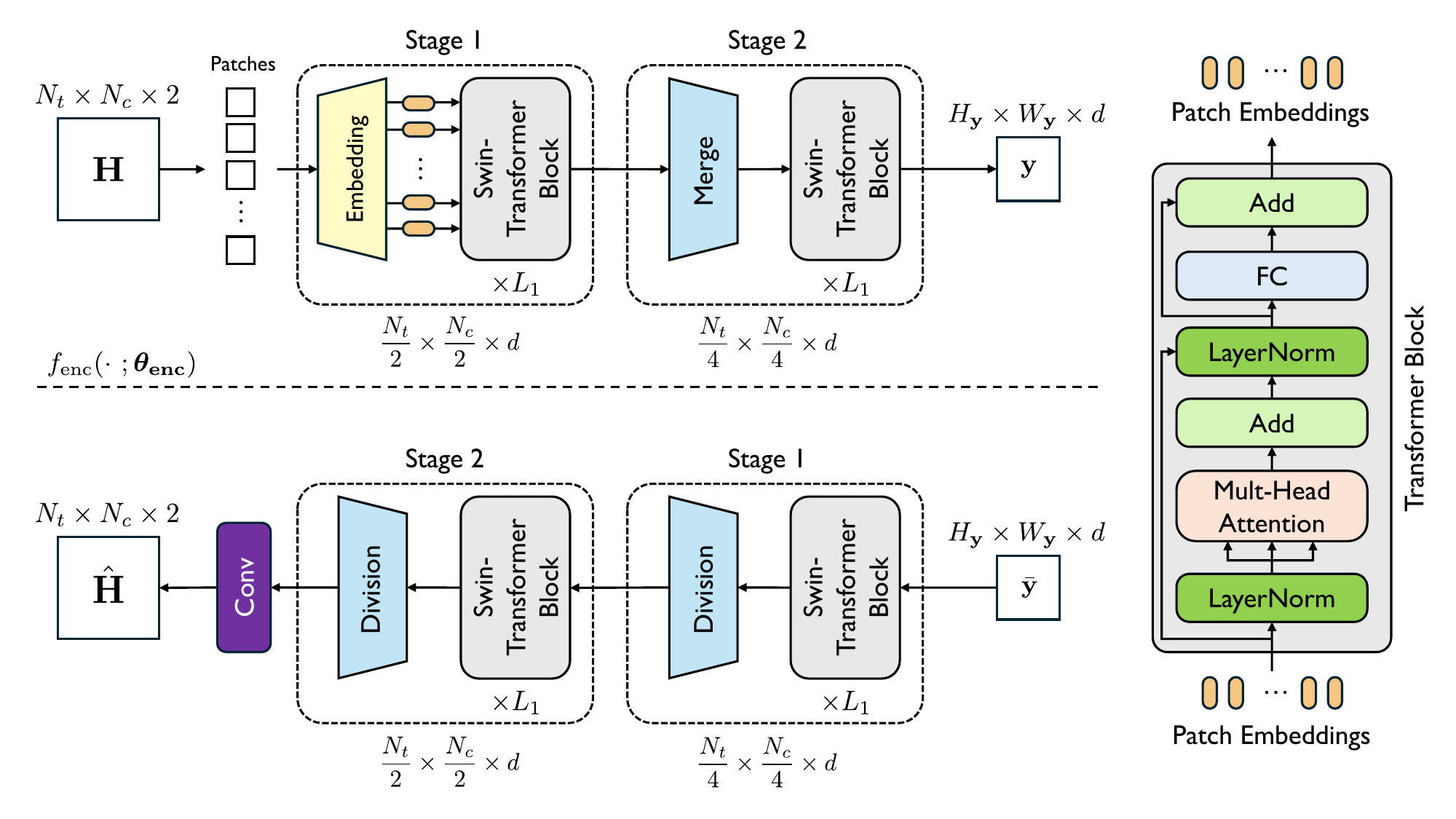}}
\caption{ Swin-Transformer-based NTC encoder and decoder structure for CSI Compression.}
\label{fig4:vit}
\end{figure*}
\subsection{DL-based bit-wise compression and reconstruction}

DL-based encoder $f_{\mathrm{enc}}(\cdot~;\boldsymbol{\theta}_{\mathrm{enc}})$ is a neural network with parameter $\boldsymbol{\theta}_{\mathrm{enc}}$ that compresses the preprocessed ${\bf H}$ into a feature vector ${\bf y}\in\mathbb{R}^{C\times H_{{\bf y}}\times W_{{\bf y}}}$. As  ${\bf y}$ is a continuous random variable, representing it with finite number of bits requires quantizing ${\bf y}$, denoted as ${\bf \bar y}$. If the probability mass function (PMF) of ${\bf \bar y}$, $P_{{\bf \bar y}}$, is known, a source coding technique such as arithmetic coding can compress $\bar{y}$ into a bit sequence $\brm{s}$ without distortion. Let $\ell(\cdot)$ be a function that returns the length of a bit sequence, according to Shannon's source coding theorem \cite{shannon1948mathematical}, the average length of $\brm{s}$, $\mathbb{E}(\ell(\brm{s}))$ cannot be smaller than the entropy of ${\bf \bar y}$. 

In this paper, we assume that the signal ${\bf s}$ is transmitted to the base station (BS) without any errors, thanks to forward error correction codes. At the BS, the channel is reconstructed as ${\bf \hat H}$) through a reverse process that includes compression, source decoding, and deep learning (DL)-based decoding. First, the received signal ${\bf s}$ is used to obtain ${\bf \bar y}$ through source decoding. The DL-based decoder $f_{\rm dec}$ is a neural network with trainable parameters ${\bf \theta}_{\rm dec}$ that reconstructs the channel from the quantized feature vector ${\bf \bar y}$.

The objective of encoder and decoder is to minimize distortion between the original channel ${\bf H}$ and the reconstructed channel ${\bf \hat H}$, subject to a rate constraint imposed by the feedback channel capacity $C_{\sf f}$,
\begin{align}
    &\underset{\boldsymbol{\theta}_{\mathrm{enc}}, \boldsymbol{\theta}_{\mathrm{dec}}}{\arg\min}~\mathbb{E}\left[\|{\bf H}-f_{\mathrm{dec}}(f_{\mathrm{enc}}({\bf H};\boldsymbol{\theta}_{\mathrm{enc}});\boldsymbol{\theta}_{\mathrm{dec}})\|^2_{\mathrm{F}}\right] \\
    &{\rm such~that}~\mathbb{E}[\ell(\brm{s})] \le C_{\sf f}.
\end{align}

\section{CSI Compression Via Transformer-NTC}
In this section, we present CSI compression method using transformer-based NTC technique, which is illustrated in Fig.~\ref{fig1:ntc}.

\subsection{NTC}

NTC includes an encoder at the user end, a decoder at the base station (BS), and a shared entropy model between the user and the BS. The encoder, which is a neural network, processes the image to extract its features. The entropy model learns parameters to determine the probability density function (PDF) of these features, enabling it to estimate their entropy.

At the user end, the entropy model estimates ${\bf y}$ using a neural network $q_{{\bf y}}(\cdot;\psi)$. The estimated probability of ${\bf \bar y}$ is then sent to the source code encoder, which converts ${\bf \bar y}$ into a bit sequence ${\bf s}$. The entropy coding method referenced in \cite{balle2017end} utilizes an asymmetric numerical system \cite{duda2013asymmetric} and range coding. This approach delivers performance comparable to arithmetic coding while being simpler than Huffman coding for shorter codeword lengths.

We assume that the bit sequence ${\bf s}$ is transmitted to the BS without errors through channel coding. At the BS, the source coding decoder reconstructs ${\bf \bar y}$ from the received ${\bf s}$, and this output is then processed by the decoder. The decoder, also a neural network, estimates the original image ${\bf \hat X}$ from the discrete features.

Let $p_{{\bf \bar y}}$ represent the true probability mass function (PMF) and $q_{{\bf y}}(\cdot;\psi)$ denote the estimated PMF of ${\bf \bar y}$. The loss function for NTC is defined as follows:

 \begin{align}
    \mathcal{L}&=\mathbb{E}_{\brm{X}}\left[\|\brm{X}-\hat{\brm{X}}\|^2_{\mathrm{F}}\right] + \lambda \mathbb{E}_{{\bf \bar y}}\left[-\log q_{{\bf \bar y};\bs{\psi}}({\bf \bar y}) \right] \\
    &=\mathbb{E}_{\brm{X}}\left[\|\brm{X}-\hat{\brm{X}}\|^2_{\mathrm{F}}\right] \\ &~~~ + \lambda \mathbb{E}_{{\bf \bar y}}\left[-\log p_{{\bf \bar y}} + \log p_{{\bf \bar y}} -\log q_{{\bf \bar y};\bs{\psi}}({\bf \bar y}) \right] \nonumber \\
    &=\mathbb{E}_{\brm{X}}\left[\|\brm{X}-\hat{\brm{X}}\|^2_{\mathrm{F}}\right] + \lambda H({\bf \bar y}) + \lambda KL( p_{{\bf \bar y}}\|q_{{\bf \bar y};\brm{\psi}}),
    \label{eq:ntcloss}
\end{align}
where $H(\cdot)$ is the entropy function for a discrete random variable. The first term in (\ref{eq:ntcloss}) minimizes the error between the original image and the reconstructed image, the second term in the loss minimizes the discrete entropy of the true quantized feature vector ${\bf \bar y}$, and the last term, which is the KL divergence between the true distribution $p_{{\bf \bar y}}$ and the estimate distribution $q_{{\bf \bar y};\bs{\psi}}$, updates $\bs{\psi}$ to fit $q_{{\bf \bar y};\bs{\psi}}$ to $p_{{\bf \bar y}}$ as close as possible. Therefore, minimizing (\ref{eq:ntcloss}), optimizing ${\bf \theta_{\rm enc}}$, ${\bf \theta}_{\rm dec}$, and ${\bf \psi}$, is equivalent to training a method that takes an image as input, compresses and decompresses it into a feature vector with minimum entropy and maximum retained information while estimating the probability distribution of the feature.

\subsubsection{Distribution modeling}
As NTC aims to minimize the loss function (\ref{eq:ntcloss}), the way we model $q$ directly influences the distribution of ${\bf y}$. The simplest approach, which offers the least flexibility, is to treat all elements of ${\bf y}$ as independent, as demonstrated in \cite{balle2017end}. While some studies have improved performance by using a more complex distribution for $q$ \cite{balle2018variational}, we opted for the independent model for the sake of simplicity. We can explore the impact of more complex modeling in CSI compression in future research.

\subsubsection{Swin-Transformer encoder-decoder}\label{transformer}

The Vision Transformer (ViT) \cite{dosovitskiy2020image} is a neural network architecture that leverages an attention mechanism to learn the global relationships within images. Unlike traditional CNNs, which struggle with scalability due to their limited number of parameters, ViT offers enhanced scalability. As a result, substituting CNN-based NTC encoders and decoders \cite{bumsu2024multirate, ravula2021deep, mashhadi2020distributed} with ViT is anticipated to yield improved performance. However, ViT does come with challenges, including a high number of parameters and significant computational complexity in its attention layers, which assess the relevance of multiple image patches \cite{bi2022novel}. This complexity can make it difficult to implement ViT-based neural networks on devices with restricted memory and processing power.

The Swin Transformer \cite{liu2021swin} was developed to overcome the limitations of the ViT. By organizing image patches into groups, the Swin Transformer significantly reduces both the number of parameters and computational complexity. To counteract the decoupling effect caused by this grouping, it uses a shifting operation that interleaves the groups, which improves connectivity and minimizes performance loss. As illustrated in Fig. \ref{fig4:vit}, the Swin Transformer architecture is made up of stacked Swin Transformer blocks. Each block includes several operations: embedding or scaling the input patches, a multi-head attention layer, a residual connection layer, and a fully connected layer. Important hyperparameters for the Swin Transformer are the image patch size $P$, embedding dimension $d$, number of multi-heads $h$, and block depth $L$. In this study, we utilize the Swin Transformer architecture to create a feature extractor and channel reconstructor, optimizing the balance between feature extraction quality, parameter efficiency, and computational cost.

\subsection{Multi-level quantization system}
\begin{figure}[tbp]
\centerline{\includegraphics[width=0.45\textwidth]{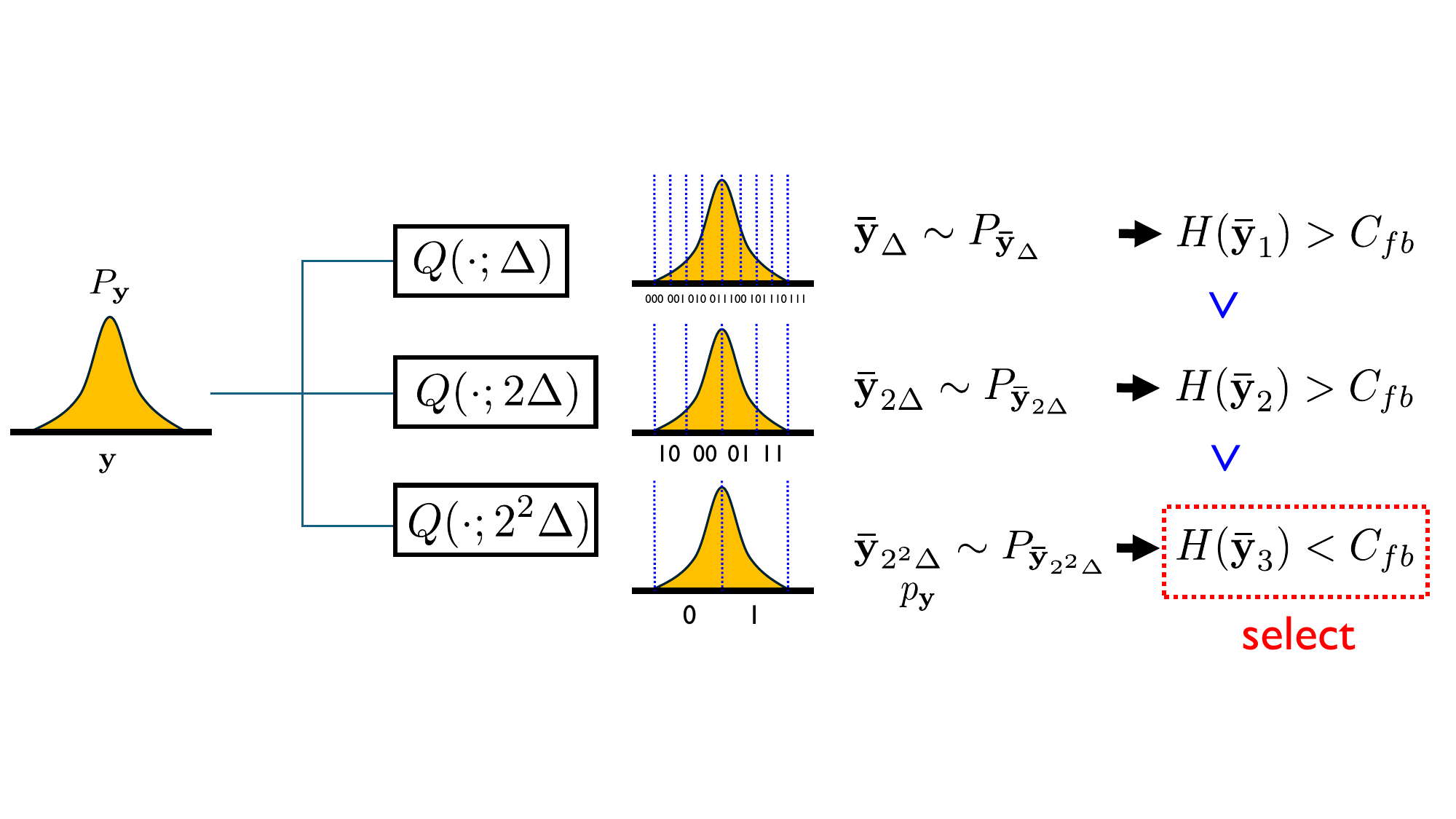}}
\caption{Illustration of multi-level uniform quantization scheme according to feedback channel condition.}
\label{fig3:multi_level}
\end{figure}

The preprocessed vector ${\bf y}$ from the NTC output is converted into a discrete vector ${\bf \bar y}$, which can then be compressed into bits using the quantization function $Q$. According to Shannon's source coding theorem \cite{shannon1948mathematical}, ${\bf \bar y}$ cannot be compressed to a rate lower than its entropy, $H({\bf \bar y})$. If the capacity of the feedback uplink channel, $C_f$, falls below $H({\bf \bar y})$ due to fluctuations in the channel, transmitting ${\bf \bar y}$ without errors becomes infeasible \cite{shannon1948mathematical}. Therefore, to compress ${\bf y}$ to a rate below $C_f$, it is crucial to lower the entropy of ${\bf \bar y}$ adaptively to a level below $C_f$, as illustrated in Fig.~\ref{fig3:multi_level}.

Vector quantization is the most effective method for quantizing ${\bf y}$. However, as the length of ${\bf y}$ increases, the computational complexity and size of the codebook grow exponentially, making it impractical to use optimal vector quantization for high-dimensional data. To address this challenge, we turn to scalar quantization, which is a more practically relevant implementation method. The non-uniform Lloyd-Max algorithm, which calculates the mean squared error (MSE)-optimal quantization regions and labels for a random variable, is one option \cite{lloyd1982quanta}. However, this algorithm is iterative and more complex than simpler methods like uniform quantization. Therefore, in our approach, we discretize ${\bf y}$ using a uniform quantization method applied consistently across all dimensions. The uniform quantizer, $Q$, maps the input to a specific point in
\begin{align}
    \left\{(n+0.5)\Delta, ~n \in \mathbb{Z}\right\},
\end{align}
which is nearest to the input.


In our previous work \cite{bumsu2024multirate}, we developed a method to modify the entropy of ${\bf \bar y}$ by adjusting the quantization level parameter $\Delta$, specifically when using a uniform quantizer for the quantization function $Q$. We proposed that the entropy of ${\bf \bar y}$ would correlate with changes in $\Delta$, and our experiments confirmed this hypothesis. Notably, if the quantized ${\bf \bar y}$ is organized into a tree structure across various regions of $\Delta$ by appropriately adjusting the bin size of the uniform quantization, then the entropy of ${\bf \bar y}$ decreases consistently as $\Delta$ increases. Given $\Delta \in \mathbb{R}^{+}$, the PMF of $\bar{y}_{\Delta}\coloneq Q(y;\Delta)$ is,

\begin{align}
    p_{n,\Delta}& \coloneq \text{Pr}\big(Q(y;\Delta)=(n+0.5)\Delta\big)\\
    &= \text{Pr}\big(n\Delta \leq y < (n+1)\Delta\big).
\end{align}
By construction, we have that
\begin{align}
    p_{n,2\Delta} 
    &= \text{Pr}\big(2n\Delta \leq y < (2n+2)\Delta\big) \\
    &= \text{Pr}\big(2n\Delta \leq y < (2n+1)\Delta\big)\\
    &\qquad +\text{Pr}\big((2n+1)\Delta \leq y < (2n+2)\Delta\big)\nonumber\\
    &=p_{2n,\Delta}+p_{2n+1,\Delta}.
\end{align}
As a result, the entropy with a smaller step size,  $Q(y;\Delta)$, is larger than that with a larger step size, $Q(y;2\Delta)$, as per \cite[Ch. 2.5]{mackay2003information}. This gives a chain of inequality:
\begin{align}
    H\big(Q(y;\Delta)\big) \geq H\big(Q(y;2\Delta)\big) \geq H\big(Q(y;2^2\Delta)\big) \geq \cdots, 
\end{align}
with $H(\cdot)$ the entropy of discrete random variable.

\section{Experiments}
\begin{figure}[tbp]
\centerline{\includegraphics[width=0.45\textwidth]{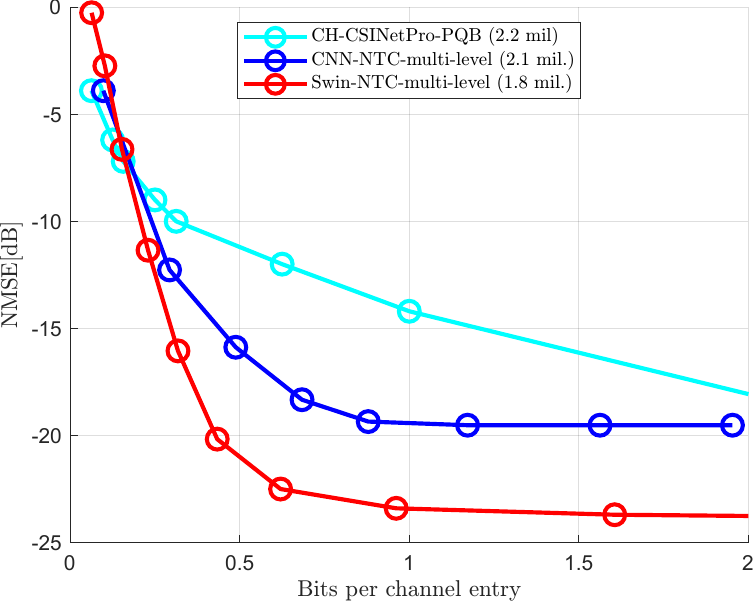}}
\caption{Rate-distortion trade-offs according to the different CSI compression techniques.}
\label{fig4:simulation}
\end{figure}


In this section, we evaluate the performance of the proposed transformer-NTC with uniform quantization parameter $\Delta$ scaling by a factor of two, and compare it with prior DL-based bit-level multi-rate compression studies. The hyper-parameters for Swin-Transformers referenced in \ref{transformer} are set as $(P,h,L)=(8,8,[6,2,2,6])$, with an embedding dimension of $d=96$ by default. For the small model with reduced transformer parameters, $d$ is set to 24. The methods for comparison are CH-CsiNet-PQB \cite{liang2022changeable} and CNN-based NTC with multi-level quantization \cite{bumsu2024multirate}. The dataset used for the simulations is the indoor 5.3 GHz dataset from COST2100 \cite{liu2012cost}, which is the most widely used for testing. The dataset's parameters are $(N_t,\tilde{N}_c,N_c)=(32,1024,32)$. We evaluate the trade-off between compression performance, measured in bits per channel, and CSI accuracy, assessed by NMSE. These metrics are defined as follows:
\begin{align}
    &\mathrm{Bits~per~channel~entry} \triangleq \mathbb{E}\left[ \frac{\ell(s)}{N_c N_t} \right]\end{align}
and 
\begin{align}
\mathrm{NMSE}             \triangleq \frac{\|{\bf H}-\hat{{\bf H}}\|^2_\mathrm{F}}{\|{\bf H}\|^2_\mathrm{F}}.
\end{align}


The simulation results shown in Fig. \ref{fig4:simulation} indicate that the CNN-NTC multi-level method, which effectively learns feature distributions, significantly outperforms CH-CsiNetPro PQB, which does not incorporate this learning. While both methods share a similar CNN structure, the key difference lies in the presence of feature distribution learning in CNN-NTC multi-level. Additionally, Swin NTC multi-level, which utilizes the Swin transformer structure, outperforms CNN-NTC multi-level. This highlights the substantial performance benefits of using the Swin transformer over a traditional CNN structure. 

CH-CSINetPro PQB shows a steady decrease in NMSE as the number of bits per channel entry increases. In contrast, NTC-based multi-level compression methods experience NMSE saturation. During training, NTC simulates the effects of quantization error by adding uniform noise to the features. Since the NTC decoder is trained to reconstruct CSI despite this noise, it remains resilient against quantization noise during inference. Even though increasing the number of bits per channel entry reduces quantization noise and enhances feature representation, the NTC decoder consistently assumes that noise is present. This assumption introduces uncertainty in feature representation, causing the NMSE performance of NTC-based methods to plateau, regardless of the increase in bits per channel entry.


\begin{figure}[tbp]
\centerline{\includegraphics[width=0.45\textwidth]{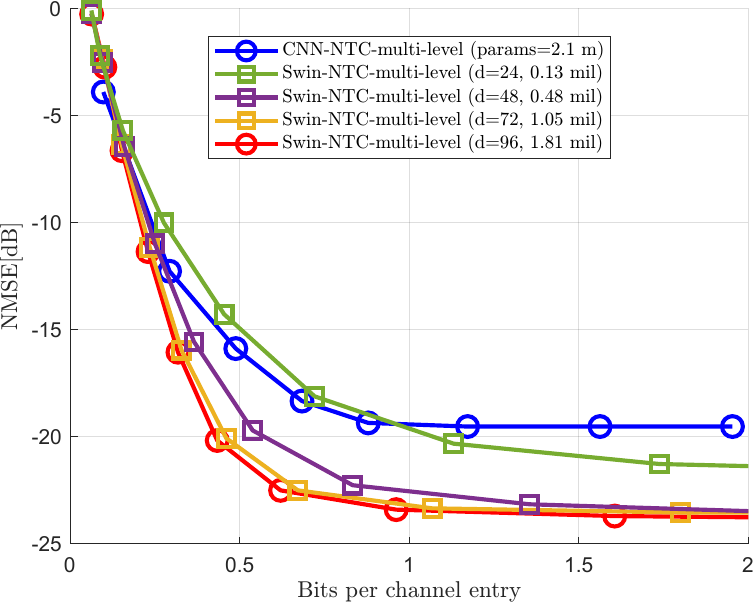}}
\caption{Rate-distortion trade-offs according to the different use of parameters.}
\label{fig5:simulation2}
\end{figure}


Fig. \ref{fig5:simulation2} illustrates the results of simulations performed under the same conditions while reducing the number of parameters in Swin-NTC-multi-level by lowering the embedding dimension $d$, which captures the implementation cost of the transformer. As shown in the figure, performance declines in proportion to $d$. Furthermore, the diminishing performance differences among the settings $d\in \{24, 48, 72, 96\}$ suggest that increasing $d$ leads to performance saturation. Notably, at ($d=24$), our method, with only 130,000 parameters—about 6\% of the parameters used in the previous state-of-the-art—outperforms the existing CNN-NTC approach. This highlights the significant advantage of using the Swin-transformer structure over CNN in terms of parameter efficiency while achieving comparable performance.

\begin{table}[tbp]
\caption{Comparison for bit-level multi-rate CSI compression methods.}
\begin{center}
\begin{tabular}{|c|c|c|c|}
\hline
\textbf{Bit-level multi-rate}&\multicolumn{3}{|c|}{\textbf{Metric}} \\
\cline{2-4} 
\textbf{CSI compression method} & \textbf{\textit{parameter}}& \textbf{\textit{\# of models}}& \textbf{\textit{Epoch}} \\
\hline
CH-CsiNet-Pro \cite{bi2022novel}& $2.2 \times 10^6$ & 5 & 2000 \\
\hline
CNN-NTC-multi-level \cite{bumsu2024multirate}& $2.1 \times 10^6$ & 1 & 200 \\
\hline
Swin-NTC-multi-level & $1.8 \times 10^6$ & 1 & 200 \\
\hline
Swin-NTC-multi-level$_{small}$ & $0.13 \times 10^6$ & 1 & 200 \\
\hline
\end{tabular}
\label{tab1}
\end{center}
\end{table}

The resources used to achieve the performance shown in Fig. \ref{fig4:simulation} are compared for each method in Table \ref{tab1}. CH-CsiNet-Pro requires the most parameters and involves training five separate models, each needing 2,000 epochs. In contrast, both CNN-NTC-multi-level and Swin-NTC-multi-level use a single model and only require 200 epochs for training.

\section{Conclusion}

This paper presents a deep learning-based image compression technique designed to compress CSI in OFDM-based massive MIMO systems. We model the frequency-domain channel matrix as a high-dimensional complex image. By utilizing a transformer-based neural network, our method optimizes both the generation of latent vectors and the distribution needed for effective entropy coding. We also introduce a multi-rate CSI compression method that incorporates a scalar nested-lattice uniform quantization scheme. Simulations demonstrate that our approach offers an excellent rate-distortion trade-off while using significantly fewer neural network parameters, showcasing its efficiency and scalability.

\section*{Acknowledgment}
This work was supported by Institute of Information \& communications Technology Planning \& Evaluation (IITP) grant funded by the Korea government (MSIT) (2021-0-00161, Post MIMO system research for massive connectivity and new wireless spectrum), and by the National Research Foundation of Korea (NRF) grant funded by the Korea government (MSIT).(No.RS-2023-00208552)

\bibliographystyle{ieeetr}
\bibliography{bib/IEEEabrv, bib/ref}

\vspace{12pt}
\end{document}